\documentclass[prl,twocolumn,showpacs,superscriptaddress]{revtex4}

\usepackage{graphicx}
\usepackage{amsmath}
\usepackage{amssymb}
\usepackage{amsthm}

\begin{document}
\bibliographystyle{apsrev}

\title{Entanglement, quantum phase transitions, and density matrix
renormalization}

\author{Tobias J.\ Osborne}
\email[]{tjo@maths.uq.edu.au}
\affiliation{Department of Mathematics, University of Queensland 4072,
Australia}
\affiliation{Centre for Quantum Computer Technology, University of
Queensland 4072, Australia}
\author{Michael A.\ Nielsen}
\email[]{nielsen@physics.uq.edu.au}
\affiliation{Centre for Quantum Computer Technology, University of
Queensland 4072, Australia}

\date{\today}

\begin{abstract} 
We investigate the role of entanglement in quantum phase transitions, 
and show that the success of the density matrix renormalization group (DMRG) 
in understanding such phase transitions is due to the way it preserves 
entanglement under renormalization.  
We provide a \emph{reinterpretation} of the DMRG
 in terms of the language and tools 
of quantum information science which allows us
to rederive the DMRG in a physically transparent way.
Motivated by our reinterpretation we suggest a modification of the DMRG 
which manifestly takes account of the 
entanglement in a quantum system. This modified renormalization scheme is shown,
in certain special cases, to preserve more entanglement 
in a quantum system than traditional 
numerical renormalization methods. 
\end{abstract}


\maketitle

A fundamental difference between quantum and classical physics is the
possible existence of nonclassical correlations between distinct
quantum systems.  The physical property responsible for the
nonclassical correlations is called {\em
entanglement}~\cite{schrodinger:1935,bell:1964,bennett:1996}.
Recently~\cite{aharonov:1999,nielsen:1998,preskill:2000}, it has been
conjectured that entanglement plays a critical role in the quantum
phase transitions which occur in interacting quantum lattice
systems~\cite{sachdev:1999} at zero temperature.

Unlike ordinary phase transitions, quantum phase transitions take
place at zero temperature as one or more physical parameters of the
system are varied.  At the critical point there are long-range
correlations in the system, as in a conventional phase transition.
However, because the system is at zero temperature, and assuming there
is no ground-state degeneracy, the system must be in a pure state.  It
follows that the correlations which are the principal experimental
signature of a quantum phase transition are due to long-range
entanglement in the ground state of the system.

A variety of methods have been developed to understand quantum phase
transitions.  One of the most successful is the density matrix
renormalization group (DMRG)~\cite{noack:1999}, a numerical technique
for finding accurate approximations to the ground and low-lying
excited states of interacting quantum lattice systems.  The DMRG was
developed in response to the poor results obtained from Wilson's
numerical renormalization group (RG)~\cite{wilson:1975} when applied
to general quantum lattice systems.  White~\cite{white:1992}
identified the mathematical origin of the breakdown of Wilson's RG as
the failure of the method to properly take into account the effect of
boundary conditions during iterations.

The focus of this paper is not on proving technical results.  Rather,
the purpose is to provide a novel physical picture of the DMRG and
quantum phase transitions based on the theory of entanglement
developed by quantum information
science~\cite{nielsen:2000a,preskillnotes}.  This physical picture
provides a fruitful way of interpreting the properties of
condensed-matter systems, substantially simplifies and physically
motivates existing derivations of properties of the DMRG, and
motivates us to propose renormalization schemes extending the DMRG. We
will explain the physical origin of the breakdown of Wilson's RG as a
failure to take into account the effects of quantum entanglement in
the ground state of the system, and show that the DMRG can be viewed
as a natural extension of Wilson's RG to take into account the
contribution entanglement makes to correlations in the ground state.
Unfortunately, owing to the limited results available on entanglement 
measures, our proposed modifications to the DMRG cannot be derived in 
closed form.  For this reason we are forced to implement the modified 
renormalization scheme numerically for idealised example systems 
with only small numbers of spins.

The structure of the paper is as follows.  First, we explain the
critical role entanglement plays in any long-range correlations
present in a quantum system at zero temperature.  Second, the
physical origin of the failure of Wilson's RG is identified.  The
relationship between the DMRG and its treatment of entanglement is
then elucidated. Finally, a numerical renormalization
scheme is presented which has the property of preserving more
entanglement in a quantum system than the DMRG.  The differences
in the entanglement preserved by the various renormalization
schemes is illustrated using the 1D Heisenberg model.

We now consider the physical origin of the correlations which occur in
systems exhibiting a quantum phase transition. For concreteness, we
restrict attention to a lattice of spin-$\frac12$ particles.  If the
ground state were a product state, then a simple calculation shows
that the spin-spin correlation function
$\langle\sigma_i^\alpha\sigma^\beta_j \rangle - \langle\sigma_i^\alpha
\rangle \langle\sigma_j^\beta \rangle $ is identically zero.  Thus, if
the correlation function is non-zero then the ground state must be
entangled, and large values of the correlation function imply a highly
entangled ground state.  It is well known that for general quantum
lattice systems the correlation function decays exponentially as a
function of the separation $|i-j|$ when the system is far from
criticality~\cite{sachdev:1999}.  When the system is at a critical
point, the correlations decay only as a polynomial function of the
separation.  At this point a fundamental change in the ground state
has occurred.  The physical property responsible for these quantum
correlations is entanglement, so that for systems approaching a
critical point the structure of the entanglement in the ground state
must undergo a transition.  Evidence for such a transition has been
obtained for the transverse Ising model \cite{osborne:2002}.

The standard technique used to derive the polynomial decay of
correlations at the critical point is the \emph{renormalization
group}~\cite{wilson:1975}.  Renormalization group techniques as
applied to quantum field theory and quantum lattice systems are
now well developed (see, for example,
\cite{peskin:1995,fisher:1998}, for reviews).  The renormalization
group works by successively discarding degrees of freedom in a
system until the dominant terms may be identified.  Each
renormalization scheme makes a decision about what degrees of
freedom in a system are important and this decision, in turn,
dictates which class of systems the method is best suited to.
Prior to 1992 the success of the numerical implementation of the
renormalization group for quantum systems was limited to the Kondo
problem~\cite{wilson:1975}, and attempts to apply Wilson's RG to
other quantum lattice systems suffered from poor convergence and
low accuracy.

Briefly, Wilson's numerical RG procedure works by isolating a portion
$A$ of a quantum lattice system containing $L$ sites and exactly
diagonalising the Hamiltonian $H_L$ of the system restricted to these
sites.  The subspace $W$ of the $m$ lowest eigenvalues and
eigenvectors is obtained.  (Note that $m$ is a parameter we choose.
It quantifies how many degrees of freedom we keep when renormalizing.)
For example, for system $A$ we might pick two adjacent spins on the
lattice, and then determine the two-dimensional subspace $W$
corresponding to the two lowest eigenvalues of $H_L$.  Once this is
done all operators are ``renormalized'' (effectively, averaging out
degrees of freedom thought to be unimportant), by projecting onto the
$m$-dimensional subspace $W$.  The allows us to renormalize the entire
lattice, replacing the two spins originally in $A$ with a single
renormalized spin.  On the new renormalized lattice a site is added to
$A$ and the interactions with the new site are taken into account by
constructing a new Hamiltonian $H_{L+1}$.  The exact diagonalisation
is performed again, keeping the $m$ lowest eigenvalues and
eigenvectors and the subspace $W$ is updated.  The expectation value
of any operator of interest is calculated with respect to the ground
state of $A$ (which is the approximation to the overall ground state),
and as the procedure is repeated the change in these expectation
values are noted.  Convergence is deemed to have been obtained when
this difference becomes smaller than some prespecified value.

Wilson's RG procedure was expected to work because it was assumed that
the only degrees of freedom contributing significantly to the overall
ground state would be the ground state and low-energy excitations of
the lattice portion $A$.  This assumption turns out not to be the
case, and Wilson's RG works poorly for most systems.  White
\cite{white:1992} gave a mathematical explanation of why this is the
case, in terms of the failure of Wilson's RG to properly take into
account boundary conditions on the system being renormalized. Using
the tools of quantum information science, we can provide another way
to interpret the failure of Wilson's RG.  The reason the assumption
fails is that the ground state of an interacting lattice is typically
entangled, and thus a subsystem of the total lattice cannot be
assigned a definite state.  Indeed, Wilson's RG ignores the terms
coupling $A$ to the remainder of the lattice, which are the very terms
responsible for the entanglement in the ground state causing the
correlations which signify a quantum phase transition.

The DMRG improves upon Wilson's scheme by taking into account the
entanglement between system $A$ and the remainder of the lattice.  It
does this by introducing a second subsystem of the lattice, $B$ (the
{\em environment}), in addition to the portion $A$ (now referred to as
the {\em principal system}) isolated by Wilson's RG procedure.  For
example, if the principal system $A$ is two adjacent spins in a
one-dimensional lattice, then $B$ might consist of two spins, one on
either side of $A$.  $A$ and $B$ together are referred to as the {\em
superblock}.  Using the physical picture based upon entanglement, the
DMRG may be described as follows.  First, construct the superblock
Hamiltonian and find its ground state.  Of course, this ground state
does not properly describe the entanglement between the superblock and
the remainder of the lattice.  However, provided $B$ is chosen
appropriately, most of the entanglement between $A$ and the remainder
of the lattice will in fact be entanglement between $A$ and $B$.
Physically, we expect this to happen if $B$ contains all those sites
to which $A$ most strongly couples.  The second step is to renormalize
$A$ by projecting onto those degrees of freedom of $A$ which
contribute most to the entangled ground state of the superblock, in a
sense made more precise below.  That is, we choose an $m$-dimensional
subspace $W$ of system $A$, and renormalize the operators for $A$ in a
similar manner to that described for Wilson's RG.  The difference
between the two approaches is in how $W$ is chosen: in Wilson's RG,
$W$ is chosen without taking account of ground state entanglement,
while in the DMRG ground state entanglement is critical.  This method
is iterated until convergence (using the same criteria as for Wilson's
RG) is obtained.

Motivated by this physical picture, we can now explain the mathematics
of the DMRG. Our description is equivalent to
White's~\cite{white:1992}, but is presented using techniques of
quantum information science which simplify the proof, and highlight
the key physical role entanglement plays in the DMRG.  The central
concept we need is the \emph{Schmidt
decomposition}~\cite{nielsen:2000a}, one of the main tools used in the
study of entangled states.  The Schmidt decomposition states that a
pure state $|\Psi\rangle_{AB}$ of a {\em bipartite} system with
components $A$ and $B$ may be written as a single sum over positive
real co-efficients $\sqrt{p_{\alpha}}$, \begin{equation}\label{eq:sd}
|\Psi\rangle_{AB} = \sum_\alpha
\sqrt{p_\alpha}|u^\alpha\rangle|v^\alpha\rangle, \end{equation} where
$|u^\alpha\rangle$ and $|v^\alpha\rangle$ are orthonormal eigenvectors
of the reduced density matrices $\rho_A$ and $\rho_B$, respectively.
It is easy to show that $|\Psi\rangle$ is entangled if and only if
there is more than one term in Eq.~(\ref{eq:sd}).

Which degrees of freedom should we keep to properly take into account
the entanglement between $A$ and $B$?  One reasonable choice is
provided by the DMRG projection $P$, which is defined to be the
$m$-dimensional projection on system $A$ that minimizes the functional
\begin{equation}\label{eq:dmrgp} \mathcal{S}(P) =
\left||\Psi\rangle_{AB}-P|\Psi\rangle_{AB}\right|^2.  \end{equation}
The motivation for introducing this functional is to isolate the
degrees of freedom which, if discarded, would change the state the
least.  To determine the projection which minimizes this functional we
write the state $|\Psi\rangle$ of the superblock in Schmidt form,
$|\Psi\rangle = \sum_\alpha \sqrt{p_\alpha} |u^{\alpha}\rangle
|v^{\alpha}\rangle$.  Simple algebra shows that \begin{equation}
\left||\Psi\rangle_{AB}-P|\Psi\rangle_{AB}\right|^2 =
1-\mbox{tr}\left(P\sum_\alpha p_{\alpha} |u^\alpha\rangle \langle
u^{\alpha}| \right).  \end{equation} Thus minimizing the functional in
Eq.~(\ref{eq:dmrgp}) is equivalent to maximizing
$\mbox{tr}\left(P\sum_\alpha p_{\alpha} |u^\alpha\rangle \langle
u^{\alpha}| \right)$.  The Ky Fan maximum principle~\cite{bhatia:1997}
tells us that the maximum in this expression is achieved by choosing
$P$ to be the projector onto the space spanned by the
$|u^{\alpha}\rangle$ with the $m$ largest values $p_{\alpha}$.

Summarizing, the physical origin for the failure of Wilson's RG and
the success of the DMRG lies in the way each method deals with
entanglement.  Typically the ground state of a lattice system exhibits
quantum correlations.  In order to represent the state of an isolated
portion of lattice sites $A$ it is necessary to take into account the
correlations between $A$ and the rest of the lattice $C$.  Due to the
nonlocal nature of the entanglement responsible for the correlations,
it is impossible to represent the state of $AC$ using only the local
degrees of freedom of $A$.  Thus Wilson's RG cannot represent the
correlations of $A$ with $C$ unless a full exact diagonalisation is
performed.  The DMRG approximates the correlations of $A$ with $C$ by
introducing a subsystem $B$ of $C$ in such a way that the entanglement
between $A$ and $B$ approximates the entanglement between $A$ and $C$.

We now outline another technique to determine the best degrees of
freedom to retain during the renormalization step that is easily
generalized to mixed states.  The best degrees of freedom to
retain may be thought of as identifying which $m$-dimensional
subspace will best preserve the overlap, or \emph{fidelity}, of
the original state with a state that has support only on the
smaller subspace.  We write the renormalized state in the
$m$-dimensional subspace in Schmidt form
$|\tilde{\Psi}\rangle=\sum_{\beta=1}^m\sqrt{q_\beta}|w^\beta\rangle
|x^\beta\rangle.$ The quantity we wish to maximize is the fidelity
between the renormalized and unrenormalized states,
$\langle\tilde{\Psi}|\Psi\rangle$.  This problem has been
considered previously~\cite{vidal:2000,barnum:1999} where it was
shown that the maximum fidelity occurs when the Schmidt basis for
$|\tilde \Psi\rangle$ corresponds to the Schmidt basis elements of
$|\Psi\rangle$ with the $m$ largest Schmidt coefficients.  This
derivation is therefore equivalent to the projection method
because the same subspaces are preserved.  The subspace-fidelity
derivation has the added benefit of being physically transparent
as well as being easily generalized to mixed states.  Indeed, when
the state of the superblock is mixed, that is, at finite
temperature, the projection method may be used to
show~\cite{noack:1999} that the optimal degrees of freedom to
retain in the renormalization step are the $m$ eigenvectors of the
reduced density matrix for $A$ with largest eigenvalues.  This
result may be verified with a subspace-fidelity argument where the
expression for the fidelity of a pair of mixed density matrices is
$F(\rho_1,\rho_2)=\mbox{tr}\left(\sqrt{\sqrt{\rho_1}\rho_2\sqrt{\rho_1}}
\right)^2$.

The DMRG is an excellent numerical scheme for many applications
because the state after the projection is usually close to the
original.  However, there is strong evidence that the DMRG does
not reproduce the algebraic decay of correlations with respect to
separation at the critical point~\cite{rommer:1998}, which is the
standard signature of a quantum phase transition.  
We conjecture
that the origin of this failure is due to the fact that while the
DMRG does preserve {\em some} of the entanglement between $A$ and
the remainder of the lattice, it does not preserve the {\em
maximal} amount of entanglement. To make this intuition more
precise, we need a good quantitative measure of how much
entanglement is present in a quantum state.  For pure states such
a measure of entanglement between $A$ and $B$ is the von~Neumann
entropy of the individual systems $A$ and $B$ alone,
$S(A)=-\mbox{tr}\rho_A\log\rho_A = S(B)$~\cite{bennett:1996}. We
conjecture that the reason the DMRG does not reproduce the
algebraic decay of correlations at the critical point is because
it minimizes the difference~(\ref{eq:dmrgp}) rather that the
difference between the entanglement of $AB$ and the entanglement
of $AB$ after the projection.

This conjecture motivates the construction of a projection
preserving more entanglement between the system and environment
than the DMRG. For the case where the superblock is in a pure
state we wish to determine which projection $P$ on $A$ when
applied to the joint state $|\Psi\rangle_{AB}$ of $AB$ will
preserve the maximal amount of entanglement between $A$ and $B$.
This is equivalent to maximizing
\begin{equation}\label{eq:empp}
\mathcal{F}=S(P\rho_AP/\mbox{tr}(P\rho_A)),
\end{equation}
over all $m$-dimensional projections $P$ on $A$.  To do this
maximization numerically, we introduce a basis $\{|\phi_i\rangle\}$
for $A$ and write $P = \sum_i|\phi_i\rangle \langle\phi_i|$ where the
sum runs over some subset of the basis.  The projection $P$ is
determined by varying the directions of the basis vectors until the
quantity in~(\ref{eq:empp}) is maximized.  Numerically this is done by
introducing a Lagrange multiplier for each basis vector to enforce
orthonormality.  We refer to the projection $P$ constructed by
the procedure outlined here as the entanglement maximization
projection (EMP).

For nonzero temperatures the state of the superblock $\rho_{AB}$ is
mixed. The derivation of the EMP outlined in the previous paragraph
requires a measure of the entanglement of $A$ with $B$ for mixed
states in order to maximize (\ref{eq:empp}).  For our calculations we
use the best-understood measure of mixed-state entanglement, the
\emph{entanglement of formation} $E(A:B)$~\cite{bennett:1996} of the
joint state $\rho_{AB}$.  The construction of the EMP is otherwise is
unchanged.  Unfortunately, no simple formula for the entanglement
of formation has yet been obtained.  For this reason we are only able to 
calculate the EMP for systems with small numbers of spins.  When
a simple formula for a good mixed-state entanglement measure is obtained
it may be possible to derive a general formula for the EMP. 

The EMP has the property that entanglement in the system is
preserved maximally.  The EMP does not minimize the functional
Eq.~(\ref{eq:dmrgp}) so that, in general, the state is quite
different after the projection. For this reason the EMP is not
appropriate for most calculations.  However, there is one
situation where we expect that the EMP may be useful, that is, for
calculating properties of a system at a quantum critical point. At
this point the correlations, which are due to entanglement, are
strongest.  We conjecture that the EMP (unlike the DMRG) may be
able to reproduce the algebraic decay of correlations at the
critical point, but have not yet been able to verify this
analytically or numerically.

\begin{figure}
\includegraphics{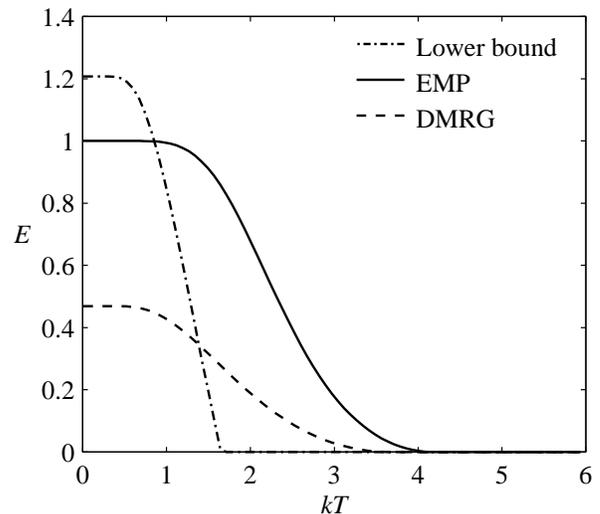}
\caption{Sites $1,2$ in the $4$-site Heisenberg model were
renormalized to a single spin via the DMRG and EMP projections
(similarly for sites $3,4$) and the entanglement between the two
renormalized spins plotted.   Also shown is a lower bound on the
actual entanglement present between $A$ and $B$ obtained from the
entropy-entanglement inequality (chapter~$8$ in
\cite{nielsen:1998}, and \cite{horodecki:2001c}). The lower bound
is expected to be a good approximation only when $kT < 1$, where
we see that the EMP provides a much better approximation to the
entanglement present in the system than does the
DMRG.}\label{fig:hsp}
\end{figure}

The difference between the entanglement preserved by the EMP and
the DMRG projection may be seen using the example of the $1$D
antiferromagnetic Heisenberg model $H = J/2\sum_{\langle i,j
\rangle}\boldsymbol{\sigma}^i\cdot\boldsymbol{\sigma}^j$
(See~\cite{nielsen:1998,wang:2001a,gunlycke:2001,arnesen:2001,meyer:2001}
for other computations of the entanglement in condensed matter
systems.) As we noted earlier, for computational ease we chose the model 
to be on
four sites.  The EMP was calculated by optimizing over the
entanglement of formation $E(A:B)$ of the projected system using
the formula of Wootters~\cite{wootters:1998}.  The entanglement of
formation as a function of temperature is shown in
Fig.~\ref{fig:hsp}.  It is clear from Fig.~\ref{fig:hsp} that the
EMP better preserves the entanglement present in the system than
does the DMRG, as expected. Similar results have been obtained for
the Ising model in a transverse field, and we expect that this is
typical of strongly interacting quantum lattice systems.

In this paper we have shown that entanglement plays a central role in
quantum phase transitions.  We have shown that the success of the DMRG
--- one of the principal numerical techniques used to solve quantum
lattice systems near criticality --- is due to the way it takes
account of the entanglement in such systems.  We have also proposed a
modification of the DMRG which preserves more entanglement during
the renormalization step, and which we conjecture may be capable
of reproducing the polynomial decay of correlations near the
critical point.  
We believe our results show that the techniques
of quantum information science can be fruitfully applied to obtain
fundamental insights into the properties of many-body systems.

\begin{acknowledgments} We thank Dorit Aharonov and John Preskill for
stimulating and encouraging discussions about entanglement and phase
transitions, and Jennifer Dodd for helpful conversations which
substantially clarified the manuscript.  This work was supported by an
Australian Postgraduate Award to TJO.  \end{acknowledgments}

\bibliography{dmrg}

\end{document}